\documentclass[12pt]{article}
\usepackage{epsfig}
\usepackage{amssymb}
\usepackage{amsmath}
\usepackage{bm} 
\usepackage{comment}
\renewenvironment{thebibliography}[1]
	{\begin{list}{[\arabic{enumi}]}
	{\usecounter{enumi}\setlength{\parsep}{0pt}
\setlength{\leftmargin 0.52cm}{\rightmargin 0pt}
	 \setlength{\itemsep}{6pt} \settowidth
	{\labelwidth}{#1.}\sloppy}}{\end{list}}
\begin{document}
%
%
%
%
%
\setlength{\jot}{10pt} 
%
%
\thispagestyle{empty} 
%
%
%
%
\vspace*{10mm}
\begin{center}  
\baselineskip 1.5cm 
{
\Large\bf
On Quantum States for Angular Position and Angular Momentum  of Light
}
\\[5mm]  
\normalsize 
\end{center} 
\begin{center} 
\vspace*{5mm}
{\centering 
{\large Bo-Sture K. Skagerstam\footnote{Corresponding author}$^{,}$\footnote{Email address: bo-sture.skagerstam@ntnu.no}
\vspace*{1mm}
\\
Department of Physics \\  Norwegian University of Science and Technology - NTNU \\ N-7491 Trondheim,  Norway}}
%
%
\\
%
{\large and}
\\
{\centering 
{\large Per K. Rekdal\footnote{Email address: per.k.rekdal@himolde.no}
\vspace*{1mm}
\\
Molde University College\\  PO Box 2110, N-6402 Molde,  Norway}}
%
\end{center} 
%
%
%
%
%
%
%
%
%
\vspace*{2.5mm}
\begin{abstract} 
%
%
\noindent In the present paper we construct  a properly defined quantum state expressed in terms of elliptic Jacobi theta functions for the self-adjoint observables angular position $\theta$  and the corresponding angular momentum operator $L = -id/d\theta$ in units of $\hbar =1$.    The  quantum  uncertainties  $\Delta \theta$ and  $\Delta L$ for the  state are well-defined  and are, e.g.,  shown to give a lower value of the uncertainty product $\Delta \theta \Delta L$  than the   minimal uncertainty states of Ref.\cite{Padgett_2004}. The mean value $\langle L \rangle$ of the state   is not required to be an integer. In the case of any half-integer mean value  $\langle L \rangle$  the state constructed exhibits a remarkable critical behaviour with  upper and lower bounds $\Delta \theta < \sqrt{\pi^2/3 -2}$ and $\Delta L > 1/2$.
\vspace{1mm}
\end{abstract} 
%
%
%
%
%
%
\vspace{0.5cm}
\newpage
\setcounter{page}{1}
%
%
%
\indent Motivated by the ingenious experimental procedures to generate restricted values of  angular position and angular momentum degrees of freedom (see, e.g., Refs.\cite{Padgett_2004,Padgett_2005,Barnett_2007, Padgett_2010}), we consider a description of such limitations in terms of a well-defined  pure  quantum state. Restrictions of the angular phase of degree of freedom must be accompanied by  restrictions on  angular momentum degrees of freedom of, e.g.,  light pulses (see Ref.\cite{Padgett_Phys_Today_2004} for a general presentation and Refs.\cite{OAM_2003} for a limited selected set of original papers on the notion of angular momentum for light pulses).  Experimental and theoretical explorations have been extended to the observations of fractional angular momentum  as well as to the use of single photon sources (see, e.g., Refs.\cite{Leach_2002,Leach_2004,Berry_2004,Woerdman_2005,Tao_2005, Yao_2006, Tanimura_2015, Wang_2015, Balantine_2016, Mitri_2016, Pan_2016, Deng_2019, Huang_2019,Chen_2021,Deach_2022, Wang_2022,Liang_2023}).  Here we, in particular, notice that the presence half-integer angular momentum appears to play  a very special role as emphasised  in, e.g., Ref.\cite{Balantine_2016}.  The quantum state as discussed   in  the present paper  turns also out to lead to   unique signatures   for half-integer angular momentum  mean values.

  In order to describe photon sources and the detection of angular momentum degrees of freedom  it is of importance to carry out a detailed second-quantization of the electromagnetic field making use of  appropriate normal mode functions. This has been investigated in great detail in the literature in various contexts requiring considerable efforts (see, e.g., Refs.\cite{barabosa_2000,Matula_2013} and references cited therein). At the single-photon level one predicts  fractional angular momentum in a reduced physical dimension (see, e.g., Refs.\cite{Balantine_2016,Pan_2016} and references cited therein). In the present paper we  limit ourselves to a transverse orbital angular momentum degree of freedom  for mode functions in a reduced physical propagation dimension and a construction of a possible well-defined quantum state according to the basic rules of quantum mechanics.
Elsewhere related charged quantum states have previously been discussed \cite{Skagerstam_2024} in the context of  quantum states  for charged q-bits. The results presented here can be applied to such systems as well.

\indent We  consider the following periodic extension of the minimal dispersion intelligent pure states as discussed in Ref.\cite{Padgett_2004}, later revisited  in Ref.\cite{Padgett_2005},  in terms  of a  parametric representation with  a real-valued  parameter $\lambda > 0$, i.e.,  
\begin{gather} 
\psi(\theta) =
 N \sum_{n=-\infty}^\infty  f(\theta - \bar{\theta} + 2\pi n)
  \label{eq:one_phase_1}	~ ,
\end{gather}
and where 
\begin{gather} 
f(x) = e^{\displaystyle{ix\bar{l}}} e^{\displaystyle{-\lambda x^2/2}} ~ .
\label{eq:one_phase_2}
\end{gather}
The range of $\theta$ is restricted  in terms of  an in principle arbitrary  off-set value $\theta_0$ for the phase $\theta$  such that $\theta_0  \leq \theta   \leq \theta_0 +2\pi$.    Apart from the  $\bar{\theta}$-dependence, the $n=0$ contribution in Eq.(\ref{eq:one_phase_1}) corresponds to the minimal dispersion state as first discussed in Ref.\cite{Padgett_2004}. As in this reference we find it  convenient to make use of the choice $\theta_0=- \pi$  since  this  naturally leads to  $\langle \theta \rangle = 0$  when $\bar{\theta}=0$.  Here $N$ is a real-valued normalization constant for the state $\psi(\theta)$ and $\overline{l}$ an additional real-valued parameter all of which will be discussed in more detail  later on. 

In passing we notice that a periodic extension of the form as in Eq.(\ref{eq:one_phase_1}) has a resemblance to the quantum states considered by M. Born \cite{Born_1955} in a discussion of the notion of probabilities in classical and quantum physics and the role of real numbers in physics. We will comment on these fundamental issues below where we, in particular, exhibit an exponential sensitivity to the difference between rational and real numbers for the observables under consideration.  

\indent The state $\psi(\theta)$ can  now be expressed in terms of a well-known  elliptic $\vartheta_3$-function (see, e.g.,  Refs.\cite{Mumford_1983}), i.e., 
\begin{gather} 
\psi(\theta) =N e^{ \displaystyle{i\overline{l}(\theta -\bar{\theta})}} e^{-\displaystyle{\lambda(\theta -\bar{\theta})^2/2}} 
 \vartheta_3[ \pi\left(\bar{l} +i\lambda(\theta-\bar{\theta})\right), e^{\displaystyle{-2\lambda\pi^2}}]
 \label{eq:two_phase_3} ~ ,
\end{gather}
using the definition 

\begin{equation} 
	\vartheta_3[z,q] = \sum_{n=-\infty}^\infty q^{\displaystyle{n^2}} e^{\displaystyle{2niz}} ~ ,
 \label{eq:two_phase_4} 
\end{equation}
within the unit disk $|q|< 1$.
The state $\psi(\theta)$  has  well-defined properties, i.e., it is continuous and differentiable for all values $\theta \in [-\pi,\pi]$,  including the boundary $\theta = \pm \pi$ in contrast to the state considered in Ref.\cite{Padgett_2004}.  One can actually be more precise in a mathematical sense and show that $\psi(\theta)$ belongs to   states for which  $\theta$ and  $L = -id/d\theta$ are self-adjoint operators (see, e.g., Example 1 in  Ref.\cite{Gesztesy_1978} and related discussions in Refs.\cite{Glazman_1963,Kraus_1965,Kato_1995,Geloun_2012}) as required by the fundamental rules of quantum mechanics. The representation of $\psi(\theta)$ in the form of Eq.(\ref{eq:two_phase_3}) is suitable for considerations of large values of the parameter $\lambda$.  
\newline
%
%
\begin{figure}[htb]  
\vspace{-0.5cm}
\centerline{\includegraphics[width=18cm,angle=0]{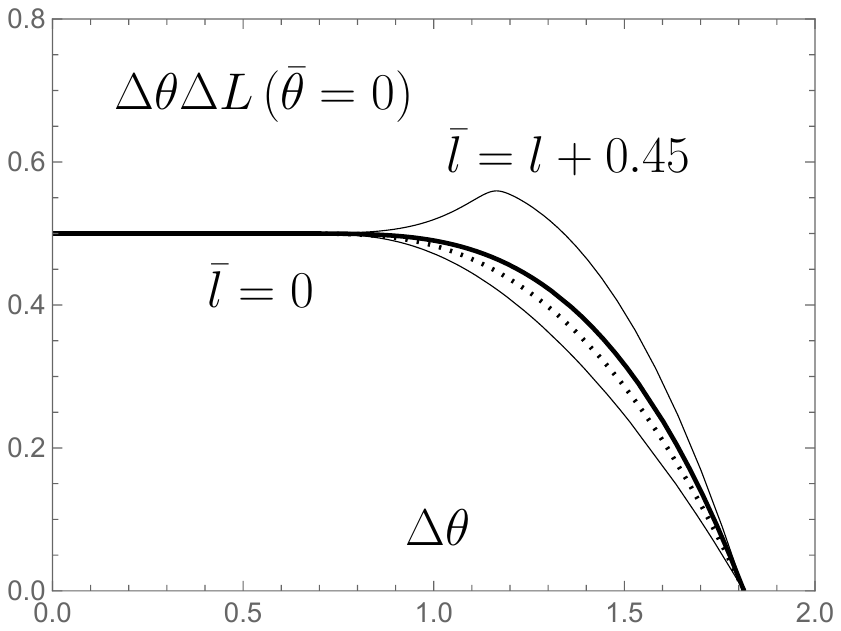}}
\vspace{-13.0cm}                 
\caption{Uncertainty products $\Delta\theta\Delta L$ as a function of the phase uncertainty $\Delta\theta$  all with the parameters $\bar{\theta}=0$  such that $\langle \theta \rangle =0$. The solid curve is the properly renormalized $n=0$ contribution in Eq.(\ref{eq:one_phase_1}) with $\bar{l}=0$  corresponding to the state of Ref.\cite{Padgett_2004}.  The dotted curve corresponds to all  terms in Eq.(\ref{eq:one_phase_1}) with $\bar{l}=0$. The lower thin line corresponds to equality in the $\theta$ and $L$  general uncertainty relation Eq.(\ref{eq:two_phase_18}) with $\bar{l}=0$.  As for a comparison the upper thin line corresponds to  Eq.(\ref{eq:one_phase_1}) with  $\bar{l}= l+ 0.45$ for any integer $l \ge 0$.
%
%
%
%
}
\label{fig:uncertainty_1}
\end{figure}
%
%
%
%
\noindent An alternative and explicit expression for $\psi(\theta)$ according to Eq.(\ref{eq:two_phase_3}) which is more adapted for a small $\lambda$ expansion can now be obtained by making use of general properties of $\vartheta_3$-functions under modular transformations (see, e.g., Refs.\cite{Mumford_1983}). For our purposes, and for the convenience of the reader, we make this explicit  in terms  of a Poisson summation technique (see, e.g., Refs.\cite{Rudin_1970,Schleich_1993})  by noticing  that  the state $\psi(\theta)$ according to Eq.(\ref{eq:one_phase_1}) is a periodic function.  It can then be  expanded in a Fourier series 
\begin{gather} 
\psi(\theta)   = \sum_{n=-\infty}^{\infty}c(n)  e^{ \displaystyle{in\theta}} 
\label{eq:two_phase_5} ~ ,
\end{gather}
such that 
\begin{gather} 
c(n) = \frac{1}{2\pi} \int_{-\pi}^{\pi} d\theta   e^{ -\displaystyle{in\theta}} \psi (\theta) = \frac{N}{2\pi} \int_{-\infty}^{\infty} d\theta  e^{\displaystyle{-in\theta }}f(\theta - \bar{\theta}) = \frac{N}{\sqrt{2\pi \lambda}}
e^{ -\displaystyle{in\bar{\theta}}} e^{ -\displaystyle{(n - \bar{l})^2/2\lambda}} 
 \label{eq:two_phase_6} ~ .
\end{gather}
%
%
\begin{figure}[htb]  
\vspace{-0.5cm}
\centerline{\includegraphics[width=18cm,angle=0]{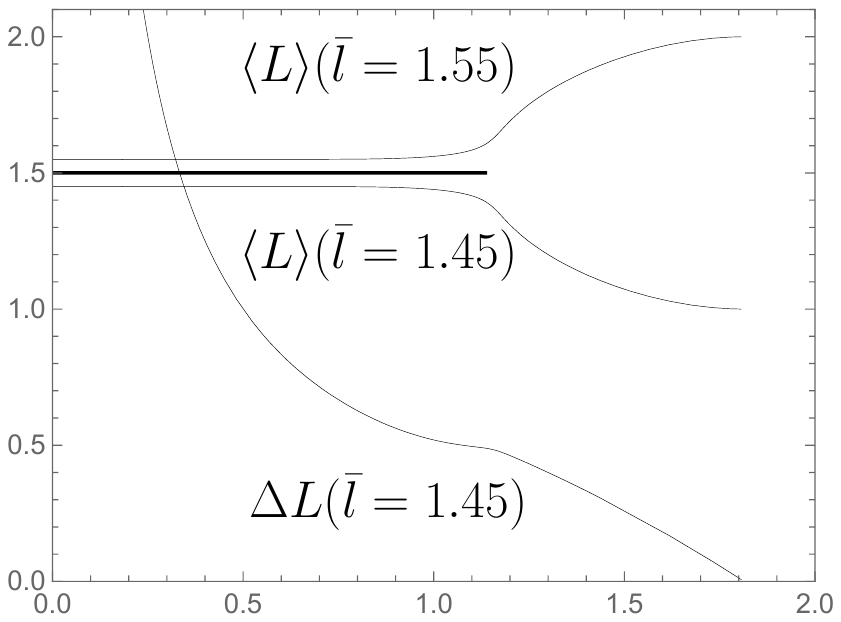}}
\vspace{-13.0cm}                 
\caption{As in Fig.\ref{fig:uncertainty_1}, the mean values $\langle L \rangle$ and the uncertainty  $\Delta L$ as a function of the uncertainty $\Delta\theta$  with $\bar{\theta} =0$ for various values of the parameter $\bar{l}$. The horizontal solid curve corresponds to a half-integer value $\langle L \rangle= 3/2$ which ends asymptotically at the exact critical value  $\Delta\theta < \sqrt{\pi^2/3 -2}$ and with the exact lower bound $\Delta L > 1/2$. The uncertainties $\Delta \theta $ and $\Delta L$ are actually is independent of the parameters $\bar{\theta}$ and  $\bar{l}= l + 1/2$ for any positive integer $l$.   The thin lines for $\langle L \rangle$, are here, and for reasons of simplicity, illustrated with $l=1$,  corresponding  in general to the values of  $\bar{l}= l + \epsilon$ with  $1/2< \epsilon < 1$ or $0 < \epsilon < 1/2$, respectively.
%
%
%
%
}
\label{fig:uncertainty_2}
\end{figure}
%
%
%
%
It then follows that 
\begin{gather} 
\psi(\theta)   = \frac{N}{\sqrt{2\pi \lambda}}\sum_{n=-\infty}^{\infty}e^{ \displaystyle{in(\theta -\bar{\theta})}}e^{ -\displaystyle{(n - \bar{l})^2/2\lambda}} \, \, ,
\label{eq:two_phase_7} 
\end{gather}
which  also can be expressed in terms of an elliptic $\vartheta_3$-function, i.e.,  
\begin{gather} 
\psi(\theta) =\frac{N}{\sqrt{2\pi\lambda}} e^{ \displaystyle{-\overline{l}^2 /2\lambda}  } \vartheta_3[ (\theta -\bar{\theta})/2 -i \overline{l}/2\lambda, e^{ \displaystyle{-1/2\lambda}}]
 \label{eq:two_phase_8} ~ ,
\end{gather}
which is rapidly converging for small values of the parameter $\lambda > 0$. 
The explicit  representation of $\psi(\theta)$ in the form as in  Eq.(\ref{eq:two_phase_7}) leads to the general normalization condition
\begin{gather} 
(\psi,\psi) \equiv \int_{-\pi}^{\pi} d\theta |\psi(\theta)|^2 = \frac{N^2}{\lambda}\sum_{n=-\infty}^{\infty}e^{ -\displaystyle{(n - \bar{l})^2/\lambda}} = N^2\sqrt{\frac{\pi}{\lambda}}\vartheta_3[ -\bar{l}\pi, e^{ \displaystyle{-\lambda \pi^2}}]=1 \, \, .
\label{eq:two_phase_9} 
\end{gather}
 The normalization factor  $N$ is therefore in general only a function of the parameters $\bar{l}$ and $\lambda$.  With the state $\psi(\theta)$ in the representation as in Eq.(\ref{eq:two_phase_7}), we can make use of Borns rule to find the properly normalized probability distribution $p(l)$ to find the  integer angular momentum $l$ component in terms of the angular momentum eigenfunction $\psi_l(\theta)= \exp(il\theta)/\sqrt{2\pi}$, i.e., 
\begin{gather} p(l) = |(\psi_l,\psi)|^2 = |\hspace{-1mm}\int_{-\pi}^{\pi }d\theta \psi_l^*(\theta)\psi(\theta)\,|^2= 
\frac{N^2}{\lambda} e^{ \displaystyle{- (l-\overline{l})^2/\lambda}} \, \, ,
\label{eq:two_phase_10} 
\end{gather}
which, of course, can be used to compute various expectation values of the observable  $L$.
The discrete nature of this Gaussian-like distribution is now such that $\lambda$ is  not directly related  to the uncertainties $\Delta \theta$ or  $\Delta L$. Below we, e.g.,  show that  $\lambda = {\cal O}(1/(\Delta \theta)^2)$ for all values of $\bar{l}$ in the large $\lambda$ limit and $\lambda = {\cal O}(1/\log(1/\Delta L))$  in the small $\lambda$ limit at least for integer values of $\langle L \rangle $.  We notice that  Eq.(\ref{eq:two_phase_7}), or making use of the probability distribution $p(l)$ above,  implies the general  and useful relations
\begin{gather}
\langle L \rangle = \frac{N^2}{\lambda} \sum_{n=-\infty}^{\infty}ne^{\displaystyle{- (n-\overline{l})^2/\lambda}}   \,\,\,\, ,\,\, \,\,
 \langle L^2 \rangle = \frac{N^2}{\lambda} \sum_{n=-\infty}^{\infty}n^2e^{\displaystyle{- (n-\overline{l})^2/\lambda}} \,\, ,
\label{eq:two_phase_11}
\end{gather}
independent of $\bar{\theta}$,  and which can be used to relate the parameter $\lambda$ to the uncertainty  $\Delta L$  at least numerically. 

Furthermore, and
with $\psi(\theta)$ in the form as given in Eq.(\ref{eq:two_phase_7}), and for any value of $\theta_0$ ,  we obtain the following general expression 
\begin{gather}
	\langle \theta \rangle = \int_{\theta_0}^{\theta_0 +2\pi }d\theta\theta|\psi(\theta)|^2 \nonumber \\ =\frac{N^2}{\lambda}  \sum_{m \neq n} \frac{\sin[(m-n)(\theta_0- \bar{\theta})]}{m-n} e^{ \displaystyle{- \left((m-\overline{l})^2 +  (n-\overline{l})^2 \right )/2\lambda }}  + (\theta_0 +\pi) ~ .
\label{eq:two_phase_12} 
\end{gather}
For any value of $\bar{l}$ we  observe  that due to Eq.(\ref{eq:two_phase_12}) the choice $\theta_0 = -\pi$   leads to  $\langle \theta \rangle = 0$ only if the  parameter $\bar{\theta} =0, \pm \pi$. 
A similar expression can be obtained for $\langle \theta^2 \rangle$ in a straightforward manner with the general result
\begin{gather}
	\langle \theta^2 \rangle  =2\frac{N^2}{\lambda}  \sum_{m \neq n} \frac{\cos[(m-n)(\bar{\theta} - \theta_0)]}{(m-n)^2} e^{ \displaystyle{- \left((m-\overline{l})^2 + (n-\overline{l})^2 \right )/2\lambda }}  \nonumber \\
+\, 2\frac{N^2}{\lambda}(\theta_0 + \pi)  \sum_{m \neq n} \frac{\sin[(m-n)(\theta_0 - \bar{\theta})]}{(m-n)} e^{ \displaystyle{- \left((m-\overline{l})^2 + (n-\overline{l})^2 \right )/2\lambda }}  + \frac{(\theta_0 + 2\pi)^3 - \theta_0^3}{6\pi} \,\, .
\label{eq:two_phase_13} 
\end{gather}
With $\theta_0 = -\pi$ we,  in particular,  have that
\begin{gather}
	\langle \theta^2 \rangle  =2\frac{N^2}{\lambda}  \sum_{m \neq n} \frac{\cos[(m-n)(\bar{\theta} - \pi)]}{(m-n)^2} e^{ \displaystyle{- \left((m-\overline{l})^2 + (n-\overline{l})^2 \right )/2\lambda }}  + \frac{\pi^2}{3} ~ .
\label{eq:two_phase_14} 
\end{gather}
In the course of the present work  we have found that less general expressions than Eqs.(\ref{eq:two_phase_9}) and (\ref{eq:two_phase_14}) have been discussed in the literature  in the context of a minimum  entrophic angular position and angular momentum quantum state \cite{Yao_2014}. There  it was also argued that a proposed quantum state could be of use in theoretical and/or experimental explorations of the notion of a phase observable in quantum physics  which we also presume can be the case for the more general quantum state Eq.(\ref{eq:two_phase_3}).

It now follows form Eqs.(\ref{eq:two_phase_12})   and  (\ref{eq:two_phase_13}) that the  implicit dependence of the parameter $\lambda$  can in general be eliminated in terms of the dispersion $\Delta \theta$ which, however,  requires numerical but straightforward considerations. If, finally,   the parameter $\bar{l}$ is any integer Eq.(\ref{eq:two_phase_7}) simplifies to
\begin{gather} 
\psi(\theta) =\frac{N}{ \sqrt{2\pi\lambda} } e^{ \displaystyle{i\bar{l}(\theta - \bar{\theta})}  } \vartheta_3[ (\theta - \bar{\theta})/2, e^{ \displaystyle{-1/2\lambda}}]
 \label{eq:two_phase_15} ~ ,
\end{gather}
and the normalization factor $N$ is determined by the relation
\begin{gather} 
\frac{N^2}{\lambda}\vartheta_3[0, e^{ \displaystyle{- 1/\lambda}}] = 1\, \, .
\label{eq:two_phase_16} 
\end{gather}
For integer values of $\bar{l}$ it also follows from Eq.(\ref{eq:two_phase_11}) that
\begin{gather}
\langle L \rangle = \bar{l} \,\,\,\, ,\,\,\,\,
 (\Delta L)^2  = \frac{N^2}{\lambda} \sum_{n=-\infty}^{\infty}n^2e^{\displaystyle{- n^2/\lambda}} \,\, .
\label{eq:two_phase_17}
\end{gather}

In the analysis of the physical properties of the state $\psi(\theta)$  we are going to make use of the  general form of the  uncertainty relation for the observables $\theta$ and $L$, i.e., the   Cauchy-Schwarz inequality, which now takes the following exact form
\begin{gather}\nonumber
(\Delta{\theta})^2(\Delta{L}^2)  \ge |((\theta - \langle \theta \rangle)\psi,(L - \langle L \rangle)\psi)|^2 \\
 =\frac{1}{4}|({\theta}\psi,{L}\psi)-({L}\psi,{\theta}\psi) |^2 + \frac{1}{4}|({\theta}\psi,{L}\psi)+({L}\psi,{\theta}\psi) -2\langle \theta \rangle \langle L \rangle|^2 ~~~,
\label{eq:two_phase_18} 
\end{gather}
where $(\Delta{\cal O})^2 = (({\cal O}- \langle{\cal O}\rangle)\psi,({\cal O}- \langle{\cal O}\rangle) \psi)$, and where the first term leads to the Kraus lower  inequality bound \cite{Kraus_1965}  for the uncertainty product, i.e.,  $\Delta{\theta}\Delta{L} \ge |1-2\pi|\psi(\pi)|^2|/2$ after a partial integration. If we, as used in Fig.\ref{fig:uncertainty_1}, consider the special case with $\bar{\theta}=0$, i.e., $\langle \theta \rangle = 0$,  we observe that $({\theta}\psi,{L}\psi)$ is purely imaginary for the state Eq.(\ref{eq:two_phase_3}). This is so since, as can be verified using Eq.(\ref{eq:two_phase_7}),  that indeed
\begin{gather}
\nonumber 
({\theta}\psi,{L}\psi) = -i\frac{N^2}{2\lambda}  \sum_{m \neq n} \cos[(m-n)\pi]) e^{ \displaystyle{- \left((m-\overline{l})^2 + (n-\overline{l})^2 \right )/2\lambda }} \\
= \frac{i}{2}(1-2\pi|\psi(\pi)|^2)~~~,
\label{eq:two_phase_19} 
\end{gather}
for all values of $\bar{l}$ in  accordance with the general uncertainty relation Eq.(\ref{eq:two_phase_18}). Therefore a strict minimal uncertainty state must satisfy the condition $\Delta{\theta}\Delta{L} = |1-2\pi|\psi(\pi)|^2|/2.$ With the parameters as used in Fig.\ref{fig:uncertainty_1} it then follows that the state in Eq.(\ref{eq:two_phase_3}) is not a minimal uncertainty state. Nevertheless, as is also illustrated in Fig.\ref{fig:uncertainty_1}, we find for the same state lower values of the uncertainty product $\Delta \theta \Delta L$  as compared to the result of Ref.\cite{Padgett_2004}.

For large values of  the parameter $\lambda$ the state $\psi(\theta)$ according to Eq.(\ref{eq:one_phase_1}) is such that the mean value of the phase $\theta$ is close to $\bar{\theta}$ due to the  $n=0$ contribution. The state $\psi(\theta)$ can then be approximated by the properly normalized expression 
\begin{gather} 
\psi(\theta) = \left(\frac{1}{2\pi(\Delta\theta)^2}\right)^{1/4}\exp \left (i\bar{l}(\theta -\bar{\theta})\right)\exp \left( - (\theta -\bar{\theta})^2/4(\Delta\theta)^2 \right)\,\, ,
\label{eq:two_phase_20} 
\end{gather}
where the range of $\theta$ can be extended to all real numbers with an exponentially small error and using the identification $\lambda = 1/2(\Delta\theta)^2$. It then follows from Eq.(\ref{eq:two_phase_20}) that $\langle L\rangle = \bar{l}$,  $\langle \theta \rangle = \bar{\theta}$,  and $\Delta\theta\Delta L = 1/2$ as well as $((\theta - \langle \theta \rangle)\psi,(L - \langle L \rangle)\psi)=i/2$ is purely imaginary. The asymptotic expression Eq.(\ref{eq:two_phase_20}) therefore corresponds to  a strict minimal dispersion state with equality in the uncertainty relation Eq.(\ref{eq:two_phase_18}) using the approximation $\psi(\pm \pi) \rightarrow \psi(\pm \infty)=0$.
As illustrated in Fig.\ref{fig:uncertainty_1},  it turns out that the approximation Eq.(\ref{eq:two_phase_20}) also well describes precise numerical considerations for a wide range of  finite values of $\lambda$. 

For sufficiently small values of $\lambda$ we proceed  without loss of generality as follows. In terms of  $\bar{l}= l + \epsilon$  with  $1/2< \epsilon < 1$ or $0 < \epsilon < 1/2$, respectively, we can make use of the small $\lambda$ expansions of the exact expressions Eqs.(\ref{eq:two_phase_9}) and (\ref{eq:two_phase_11}) in order to obtain the corresponding expressions for  $\langle L \rangle$ and $\Delta L$, i.e.,  
\begin{gather}
\langle L \rangle = l + \frac{1}{1+ e^{\displaystyle{(1-2\epsilon)/\lambda}}} \,\,\,\, , \,\,\,\, 
(\Delta L)^2 = \frac{e^{\displaystyle{(1-2\epsilon)/\lambda}}}{(1+ e^{\displaystyle{(1-2\epsilon)/\lambda}})^{\displaystyle 2}} \,\, ,
\label{eq:two_phase_21}
\end{gather}
independent of the parameter $\bar{\theta}$.
Therefore the mean value $\langle L\rangle$ in the small $\lambda$ limit approach the  integer values $l+1$ or $l$ depending on $1/2< \epsilon < 1$ or $0 < \epsilon < 1/2$, respectively, with limiting value  $\Delta L = 0$ as it should. This branching feature of the state $\psi(\theta)$ is illustrated in Fig.\ref{fig:uncertainty_2}  for the case $l=1$.  For integer values of $\langle L \rangle$, corresponding to the limiting values $\epsilon=0$ or $\epsilon =1$, we obtain from Eq.(\ref{eq:two_phase_21}) that $\lambda = 1/\log(1/(\Delta L)^2)$ in the small $\lambda$ limit.

By inspection of Eq.(\ref{eq:two_phase_7}) it,  furthermore, follows  that in the small $\lambda$ limit  the quantum state $\psi(\theta)$ takes the following form 
\begin{gather} 
\psi(\theta) = \frac{1}{\sqrt{2\pi}}
\frac{1+ e^{\displaystyle{i(\theta -\bar{\theta})}}e^{\displaystyle{-(1-2\epsilon)/2\lambda}}}{(1+ e^{\displaystyle{-(1-2\epsilon)/\lambda}})^{\displaystyle 1/2}}e^{ \displaystyle{il(\theta -\bar{\theta})}}\,\, .
\label{eq:two_phase_22}
\end{gather}
In this limit   we therefore  obtain the expected $l +1$  or $l$  angular momentum eigenstates
\begin{gather} 
\psi(\theta) = \frac{1}{\sqrt{2\pi}} e^{ \displaystyle{i(l+1)(\theta -\bar{\theta})}}\,\,\,\,  , \,\,\,\,  \mbox{or} \,\,\,\,  , \,\,\,\, 
 \frac{1}{\sqrt{2\pi}} e^{ \displaystyle{il(\theta -\bar{\theta})}} \,\, ,
\label{eq:two_phase_23}
\end{gather} 
in  the case $1/2< \epsilon < 1$  or  in  the case $1/2< \epsilon < 1$, respectively, with $\Delta L = 0$ and the limiting value of $\Delta\theta$ given by $\pi/\sqrt{3}$. 

It is now clear from Eqs.(\ref{eq:two_phase_21}) and (\ref{eq:two_phase_22}) with $\epsilon = 1/2$ that  exact half-integer  values of $\bar{l} = l + 1/2$, where $l$ is a positive integer, play a very special role, at least for sufficiently small $\lambda$.  The asymptotic form of $\psi(\theta)$  is then a superposition of the $l+1$ and $l$ angular momentum eigenstates  in Eq.(\ref{eq:two_phase_23}), i.e., 
\begin{gather} 
\psi(\theta) = \frac{1}{\sqrt{4\pi}}
\left( 
e^{\displaystyle{il(\theta -\bar{\theta})}} + e^{\displaystyle{i(l+1)(\theta -\bar{\theta})}}\right) \,\, .
\label{eq:two_phase_24}
\end{gather}
 For sufficiently small values of $\lambda$, and according to Eq.(\ref{eq:two_phase_21}) or using Eq.(\ref{eq:two_phase_24}),  we then find  that $\langle L \rangle = l+ 1/2$ and $\Delta L = 1/2$ as well as the  limiting value $(\Delta \theta)^2 = \pi^2/3 -2$.   Here we notice another unique feature in the case of  $\epsilon = 1/2$ namely that $\langle L \rangle = l+ 1/2$ is actually exact and valid for all values of $\lambda >0$.  This can be verified by making use of the exact form of $\langle L \rangle$ according to Eq.(\ref{eq:two_phase_11}) and is due a simple but remarkable cancellation of contributing terms and by making use of the exact form of the normalization constant $N$ in Eq.(\ref{eq:two_phase_9}).  

For half-integer mean values of $\langle L \rangle$, and  for $\langle \theta \rangle = 0$, it also follows that the lower limit in uncertainty relation Eq.(\ref{eq:two_phase_18}) is such that equality in  $\Delta \theta \Delta L \ge 1/2$ implies a minimum uncertainty state. This is due to the fact that in this case it can be verified $|(\psi(\theta),\psi(\theta))|=0$ for all values of $\lambda$ if $\theta= \pi$.  These remarkable features are illustrated in Figs.\ref{fig:uncertainty_1} and \ref{fig:uncertainty_2} for various values of the parameter $\bar{l}$.  One can now verify that the state $\psi(\theta)$ with the parameters as in Fig.\ref{fig:uncertainty_1} is not a minimum uncertainty state.  

It is of importance to notice that for any real value of $\bar{l}$ arbitrarily close  to a half-integer rational number we always obtain the branching to angular momentum $l$ or $l+1$ eigenstates  for sufficiently small values of $\lambda$ as illustrated in Fig.\ref{fig:uncertainty_2}. The well-defined quantum state as constructed in the present work therefore provides for an explicit  example with observables that exhibit exponential sensitivity to the difference between rational numbers and real numbers in quantum mechanics. It would, of course, be of great interest if the state $\psi(\theta)$ according to Eq.(\ref{eq:two_phase_3}) could be prepared and investigated in a concrete physical situation  revealing the fundamental role of real numbers in quantum physics \cite{Born_1955} perhaps by extending  the experimental $\bar{l}=0$ procedure of Ref.\cite{Padgett_2004}.

In summary we have studied a well-defined quantum state $\psi(\theta)$ which exhibits some remarkable properties. For $\langle L \rangle$ arbitrarily close to a half-integer we have, e.g.,  shown that  the state $\psi(\theta)$ leads to two orthogonal eigenstate of $L$ in the limit as $\Delta\theta$ approaches the conventional $\pi/\sqrt{3}$. In the case of  an exact  half-integer mean value of the angular momentum $L$ the uncertainty  $\Delta\theta$ has, furthermore,   an upper precise bound $\sqrt{\pi^2/3 -2}$  which is smaller than conventional bound $\pi/\sqrt{3}$ and $\Delta L \ge 1/2$. We have also exhibited an explicit exponential sensitivity to the difference of rational and real numbers in terms of an allowed and well-defined quantum state. 
%
%
%
%
%
%
%
%
%
%
%
%
%
%
\vspace{1cm}
\begin{center}
   {\bf \large ACKNOWLEDGMENT}
\end{center}
The research by B.-S. Skagerstam (B.-S. S.)  has been supported in by NTNU and Molde University College.  B.-S.S. is grateful to J.R. Klauder for discussions and for informing us about Ref.\cite{Geloun_2012}.
The research by P.K. Rekdal  has been supported by  Molde University College.
%
%
%
%
%
%
%
%
%
\newpage
\begin{center}
   {\bf \large REFERENCES}
\end{center}
%
%
 
%
%
%
%
%
%
%
\end{document}